\documentclass[%
  twocolumn,
  superscriptaddress,
  amsmath,amssymb,
  aps,
  pre,
  showpacs,
]{revtex4}

\usepackage{graphicx}
\usepackage{dcolumn}
\usepackage{bm,ulem,color}
\usepackage{epstopdf}
\usepackage[utf8]{inputenc}

\begin{document}

\title{Clustering of vertically constrained passive particles in homogeneous, isotropic turbulence \footnote{Version accepted for publication (postprint) on  Phys. Rev. E 91, 053002 – Published 4 May 2015}}

\author{Massimo De Pietro} 
\affiliation{Dip. di Fisica and INFN,
  Universit\`a ``Tor Vergata", Via della Ricerca Scientifica 1,
  I-00133 Roma, Italy.}  

\author{Michel A.T. van Hinsberg}
\affiliation{Department of Applied Physics, J. M. Burgerscentrum, Eindhoven
  University of Technology, 5600 MB Eindhoven, The Netherlands}

\author{Luca Biferale} \affiliation{Dip. di
  Fisica and INFN, Universit\`a ``Tor Vergata", Via della Ricerca
  Scientifica 1, I-00133 Roma, Italy.}  

\author{Herman J.H. Clercx}
\affiliation{Department of Applied Physics, J. M. Burgerscentrum, Eindhoven
  University of Technology, 5600 MB Eindhoven, The Netherlands}

\author{Prasad Perlekar} \affiliation{TIFR Centre for Interdisciplinary
  Sciences, Tata Institute of Fundamental Research, 21 Brundavan Colony, Narsingi, Hyderabad 500075, India}

\author{Federico Toschi} \affiliation{Department of Applied Physics and
  Department of Mathematics and Computer Science, Eindhoven University
  of Technology, 5600 MB Eindhoven, The Netherlands and IAC, CNR, Via
  dei Taurini 19, I-00185 Roma, Italy}

\date{\today}

\begin{abstract}
  We analyze the dynamics of small particles vertically confined, by
  means of a linear restoring force, to move within a horizontal fluid
  slab in a three-dimensional (3D) homogeneous isotropic turbulent
  velocity field.  The model that we introduce and study is possibly
  the simplest description for the dynamics of small aquatic organisms
  that, due to swimming, active regulation of their buoyancy, or any
  other mechanism, maintain themselves in a shallow horizontal layer
  below the free surface of oceans or lakes. By varying the strength
  of the restoring force, we are able to control the thickness of the
  fluid slab in which the particles can move. This allows us to
  analyze the statistical features of the system over a wide range of
  conditions going from a fully 3D incompressible flow (corresponding
  to the case of no confinement) to the extremely confined case
  corresponding to a two-dimensional slice. The background 3D turbulent velocity
  field is evolved by means of fully resolved direct numerical
  simulations. Whenever some level of vertical
  confinement is present, the particle trajectories deviate from
  that of fluid tracers and the particles experience an effectively
  compressible velocity field. Here, we have quantified the
  compressibility, the preferential concentration of the 
  particles,  and the correlation dimension by changing the strength of
  the restoring force. The main result is that there exists a particular
  value of the force constant, corresponding to a mean slab depth
  approximately equal to a few times the Kolmogorov length scale, 
  that maximizes the clustering of the particles.
\end{abstract}

\maketitle

\section{Introduction}
The problem of the distribution of inertial
particles in a turbulent flow is a crucial topic
in many different fields \cite{maxey_gravitational_1987,
  balkovsky_intermittent_2001, bec_fractal_2003, bec_heavy_2007,
  toschi2009lagrangian, bec_intermittency_2010, fouxon_distribution_2012,
  gustavsson_2014, bec_gravity_driven_2014}, for instance, modeling the interactions between small particles
carried by a turbulent flow for the study of cloud
formation in the atmosphere, the development of industrial processes,
or the study of the dynamics of plankton organisms in oceans and
lakes.
 It is known that while non-inertial particles follow
exactly the flow streamlines, and are homogeneously distributed in the
fluid volume, inertial particles lighter than the fluid tend to be
trapped inside vortices, as opposed to heavier inertial particles,
that tend to accumulate in strain-dominated regions of the flow
\cite{maxey_gravitational_1987, squires1991preferential,
  wang1993settling, bec_fractal_2003}. This preferential concentration has important
consequences in the dynamics of particles under the influence of gravity 
\cite{bec_gravity_driven_2014}, and in all the situations where
the clustering of particles may have non-trivial consequences, as in,
for example, cloud formation \cite{falkovich_acceleration_2002,
  shaw_particle-turbulence_2003,grabowski_growth_2013} or the biology
of aquatic microorganisms \cite{squires1995preferential,
  durham_turbulence_2013, perlekar2013cumulative, PhysRevLett.112.044502}.

In this paper we will study an ``idealized'' situation that could be  interesting both as
a particular case of particles falling through a weakly stratified
fluid, until they reach a buoyant equilibrium, and as a 
model of the dynamics inside plankton layers. Observations of marine ecosystems often report the
striking finding of plankton populations living confined in
horizontally extended and vertically thin layers
\cite{martin2003phytoplankton, durham2012thin}. Many different
physical and biological mechanisms such as buoyancy regulation, gyrotaxis
  \cite{durham_turbulence_2013, PhysRevLett.112.044502}, or nutrient variability
  \cite{mckiver_influence_2009} may be at the basis of the formation of such
planktonic layers, and the relevance of the different mechanisms may
vary amongst different plankton species \cite{durham2012thin}.
The spatial confinement of living populations has direct consequences on 
the total population size (carrying capacity). This is the case also when, 
independently from the particular physical mechanisms, an effective 
compressibility is produced, leading to preferential accumulation 
\cite{PhysRevLett.105.144501, Perlekar2011statistics, Nelson2012biophysical, Pigolotti2012Population_Genetics, benzi2012population_dynamics_EPJ, perlekar2013cumulative, Pigolotti201372, Kalda201456}.

In this paper, we propose and analyze a simple model meant to describe the 
effects of preferential concentration on passive small particles confined to move on a vertical slab inside a chaotic and turbulent flow.
We are not interested in the biological or structural reason leading to the confinement; we will limit ourselves to imagine that there exists a bias in the equation of motion that does not allow the particles to move freely in the vertical-direction and analyze the consequences of this fact. 
To do that,  we introduce a linear restoring force, capable of providing a
tunable confinement level for particles in a specific depth
interval. These confined particles are advected by a velocity field
obtained by a  direct numerical simulation (DNS) of
homogeneous and isotropic turbulence.
Dispersion and transport processes under real marine conditions are
usually complicated by many more phenomena which we have not included
in our model. For example, stratification due to density differences
will be important for oceans and estuarine flows (salinity and
temperature gradients) and also for lakes (temperature gradients).
Density stratified turbulent flows will change the dynamics of the
flow, the particle trajectories, and the dispersion
properties~\cite{vanaartrijk2009dispersion,vanaartrijk2010vertical}. Moreover,
simulation of real plankton dynamics should include many biological
phenomena like reproduction or nutrient cycles that are not
incorporated in this model.

The paper is structured as follows: in Section
\ref{sec:ModellingAndMethods} we describe in detail the model used for
the dynamics of the particles and the parameters of the simulation. In Section \ref{sec:Observables} the  results of
the simulations are presented.  Our main result shows that  there exists a
non-trivial correlation between the distribution of passive particles
as a function of the degree of vertical confinement and the underlying
homogeneous and isotropic turbulent flow field.

\section{Modeling and methods}
\label{sec:ModellingAndMethods}
\subsection{The flow}
\label{sec:TheFlow}
The numerical integration of the homogeneous and isotropic turbulent
velocity field is performed by means of DNS of the Navier-Stokes
equations employing a standard pseudo-spectral algorithm. The domain
size is a cube with size $L^3=(2 \pi)^3$ and a $128^3$ grid has been
used for the discretization. Periodic
boundary conditions have been applied in all three directions. The
nonlinear term is dealt with by the $2/3$-rule for dealiasing,
temporal advancement is based on the Adams-Bashforth second order
(AB2) scheme.

Energy is injected at small wave numbers in order to 
achieve a stationary state. The external forcing is such that the energy 
injected in the system is constant \cite{lamorgese2005direct}. 
All values of physical parameters that follow are 
given in units of the numerical simulation.
The viscosity, in our simulation, was
$\nu = 0.01$ and the energy dissipation rate $\epsilon
\simeq 2.39$.
The Kolmogorov 
length scale is calculated using the standard dimensional argument
$\eta = (\nu^3/\epsilon)^{1/4} \simeq 0.025 $ and the Kolmogorov time scale 
is $\tau_{\eta} \simeq 0.065 $. As to the integral scales, the rms 
velocity is $v_{rms} \simeq 1.32$ and the large-scale
eddy-turnover time is $T_L = L/v_{rms} \simeq 4.7$.

The resulting Reynolds number was $Re = v_{rms}L/\nu
\simeq 800$ (corresponding to a Taylor scale Reynolds number
$Re_{\lambda} = v_{rms} \lambda / \nu \simeq 60$). 
Let us notice that the underlying flow is only moderately turbulent. This is not considered a problem, being in the sequel mainly interested in small-scales clustering, i.e., at length scales smaller than the Kolmogorov scale, where the flow would be smooth anyway (but chaotic in time and with non trivial spatial correlations). 

In total, $N_{p}=10^{5}$ particles have been injected at time $t=0$ on
a plane of constant height $z_{0}$. The particles are randomly
and homogeneously distributed within the chosen plane. The particle
equations of motion (see Section~\ref{sec:EquationsOfMotion}) are also
advanced in time using the AB2 scheme. Both fluid and particle equations
of motion were numerically integrated for about $100$ large
eddy-turnover times $T_L$.
To ensure that the dynamics of the system is statistically stationary 
(and transient phenomena have decayed), ensemble averaging starts from time $> 4\,T_L$.

\subsection{Equations of motion for the particles}
\label{sec:EquationsOfMotion}
The particles are treated as passive (i.e., they produce no feedback on
the fluid), point-like tracers with a confining force acting along the
vertical, ${\mathbf{\hat z}}$, direction. The equations of motion are:
\begin{eqnarray}
  \label{eq:eq_motion_particles}
  \nonumber
  \frac{d {\mathbf x}(t)}{dt} &=& \mathbf{u}({\bm x},t), \\
  \mathbf{u}({\bm x},t) &=&\mathbf{v}({\bm x},t)-K(z(t)-z_0)\mathbf{\hat{z}} \, ,
\end{eqnarray}
where $\mathbf{u}$ is the velocity of the particle at time $t$ at
position ${\bm x}$, $\mathbf{v}$ is the velocity of the fluid at the
particle position, $K$ is a force constant (determining the strength
of the confinement), and $z_0$ (here $z_0=L/2$) is the center of the
vertical confinement layer.\\
Equation \eqref{eq:eq_motion_particles} must 
be understood as the simplest 
(minimal) set of equations that might mimic one of the many cases 
when small particles are constrained to move on a given layer inside an 
otherwise three-dimensional (3D) volume. The physical mechanism leading 
to this constraint can have a very different origin. Here we limit ourselves to notice 
that, for example, Eq \eqref{eq:eq_motion_particles} can be formally 
derived from the Maxey-Riley 
equations \cite{maxey1983equation}, in the case of almost neutrally 
buoyant particles in a linear density profile. If we
neglect the Basset history term and the Fax{\'e}n
corrections, we can write \cite{auton1988force}:

\begin{equation}
  \frac{d\mathbf{u}}{dt}=\beta\frac{D\mathbf{v}}{Dt}-\frac{\mathbf{u}-\mathbf{v}}{\tau_{s}}+\left(1-\beta\right)\mathbf{g}
  \, ,
  \label{eq:eq_2_par}
\end{equation}
where $\beta=\frac{3\rho_{f}}{\rho_{f}+2\rho_{p}}$ is the density
ratio (with $\rho_p$ and $\rho_f$ the particle and fluid density,
respectively), $\tau_{s}=\frac{a^{2}}{3\nu\beta}$ is the particle
relaxation time (or Stokes time), and $\mathbf{g} =
-g{\mathbf{\hat{z}}}$ is the acceleration due to gravity.
We consider a linear density profile for the flow
$\rho_f(z) \simeq \rho_0 + (d \rho_f / d z)(z-z_0)$
and use the definition of the Brunt-V{\"{a}}is{\"{a}}l{\"{a}} frequency $N$,
for writing: $|d\rho_f/dz| = \rho_0N^2/g$~\cite{gill1982} (note that the density gradient is
negative for stable stratification).
We assume that $N \ll 1$ and that $\rho_p=\rho_0$, as we release neutrally buoyant particles
at the reference depth $z_0$. With this in mind, the buoyancy term in Eq. 
\eqref{eq:eq_2_par} can be expressed as:
\begin{equation}
  \left(1-\beta\right)\mathbf{g} \simeq -\frac{2 \mathbf{g} (d \rho_f
    / d z)}{3 \rho_0} (z-z_0) = - \frac{2}{3}N^2(z-z_0){\mathbf{\hat{z}}} \, .
  \label{eq:buoy}
\end{equation}
Multiplying by $\tau_s$ and rearranging terms in Eq. 
\eqref{eq:eq_2_par} we obtain:
\begin{equation}
  \mathbf{v}-\mathbf{u} = \tau_{s} \frac{d\mathbf{u}}{dt} - \tau_{s}
  \beta \frac{D\mathbf{v}}{Dt} + K (z-z_0)\mathbf{\hat{z}} \, ,
  \label{eq:eq_velocita_completa}
\end{equation}
where $K = \frac{2}{3}N^2\tau_{s}$.\\
In the limit of small $\tau_s$,
Eq.~\eqref{eq:eq_velocita_completa} can be further
simplified. Following Ref.~\cite{maxey1983equation}, for small
inertia $D{\mathbf v}/Dt \sim d{\mathbf u}/dt$ and substituting Eq.~\eqref{eq:buoy} 
in Eq.~\eqref{eq:eq_velocita_completa} under the further hypothesis 
that $(D\mathbf{v}/Dt)/g \ll 1$ we obtain our model equation \eqref{eq:eq_motion_particles}.

Other mechanisms may support or counteract confinement. Examples are
swimming of algae or buoyancy self-regulation of cells. We assume that
it is possible to model the combined effect of such mechanisms with a
(weak) background density stratification with an effective
potential. If this is the case,
Eq.~\eqref{eq:eq_velocita_completa}, and its simplified
version Eq.~\eqref{eq:eq_motion_particles} for small Stokes numbers, stay
the same, with the only difference being a modified force constant
$K$. The resulting equation of motion~\eqref{eq:eq_motion_particles}
leads to a Gaussian-like concentration profile, as we will show later (Fig. \ref{fig:z-hist-el}). 
In the remaining part of this paper we restrict ourself to and discuss
the physical aspects of the particle distribution in confined layers.
\section{Observables and results}
\label{sec:Observables}
First of all we analyze the effects of confinement on the distribution of particles at large scales, by measuring 
the  particles probability distribution, $N_{K}(z)$,
in the $z$ direction and its standard deviation, $\sigma$, from the
middle plane (also referred to as an effective layer or slab
depth). These quantities provide indications on the effective spatial
extent and the strength of the vertical confinement. In order to
characterize the implication of particle confinement on the particle
distribution in the horizontal slab of fluid within a 3D statistically
homogeneous and isotropic turbulent flow field, we also analyze 
both the two-dimensional (2D) and 3D compressibility effects as well as the velocity
correlation integrals. 
\subsubsection{Vertical distribution}
\label{sec:met.Z-distribution}
%
%
\begin{figure}
  \includegraphics[width=\linewidth]{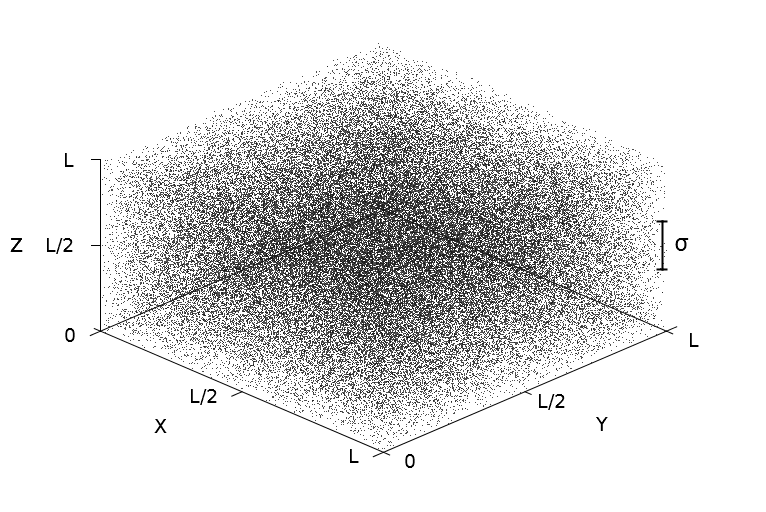}
  \vskip -0.3cm
  \includegraphics[width=\linewidth]{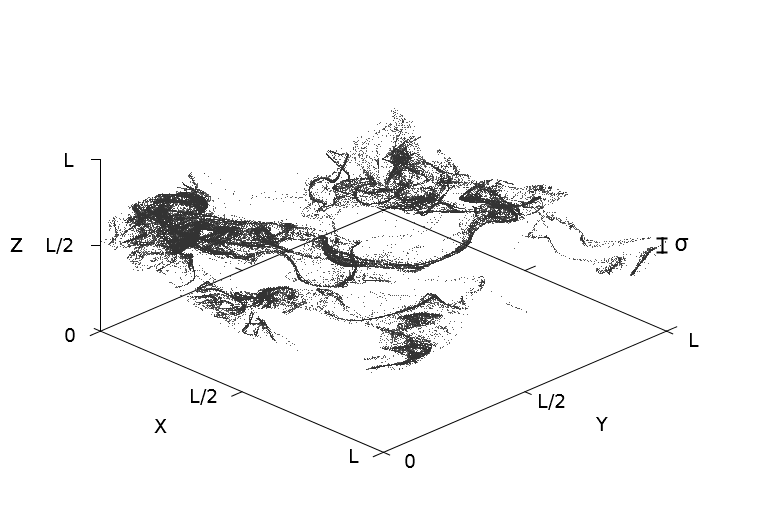}
  \vskip -0.3cm
  \includegraphics[width=\linewidth]{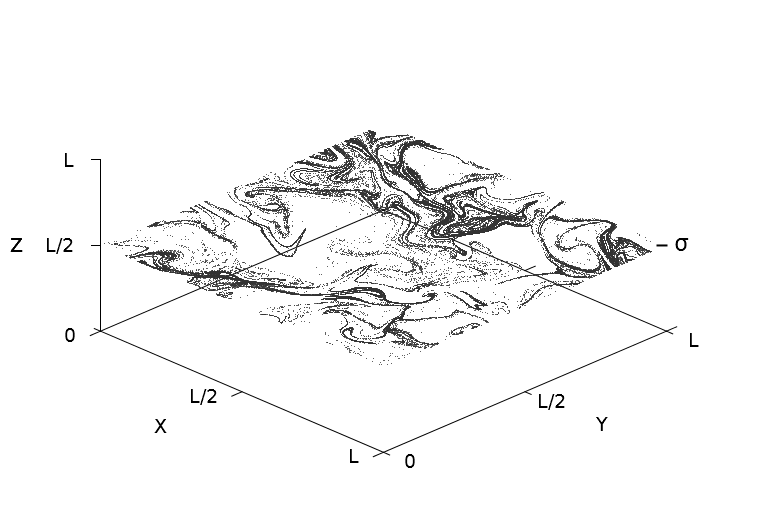}
  \caption{\label{fig:fam1-el}Plot of the spatial distribution of particles for
    different values of the force constant $K$. The magnitude of the
    slab depth $\sigma$ is also shown for comparison. (top panel) Free
    tracers case, corresponding to the choice of parameters $K=0$.
    (middle panel) Intermediate confinement case, with force
    constant $K=0.125$. The associated effective slab depth is $\sigma
    \simeq 21\eta$.  (bottom panel) Strong confinement case
    corresponding to parameters $K=6$ and $\sigma \simeq
    0.46\eta$. Here, particles are almost perfectly confined in a
    plane. The presence of the vertical confinement induces an evident
    and strong preferential concentration in the horizontal plane.}
  \label{fig:sist-elastic-plot}
\end{figure}
In Figure \ref{fig:sist-elastic-plot} we show some representative
snapshots of the system, for different values of the force constant
$K$. From these pictures it is evident that confinement has strong effects both at large and small scales.

\begin{figure}[!t]
  \includegraphics[width=\linewidth]{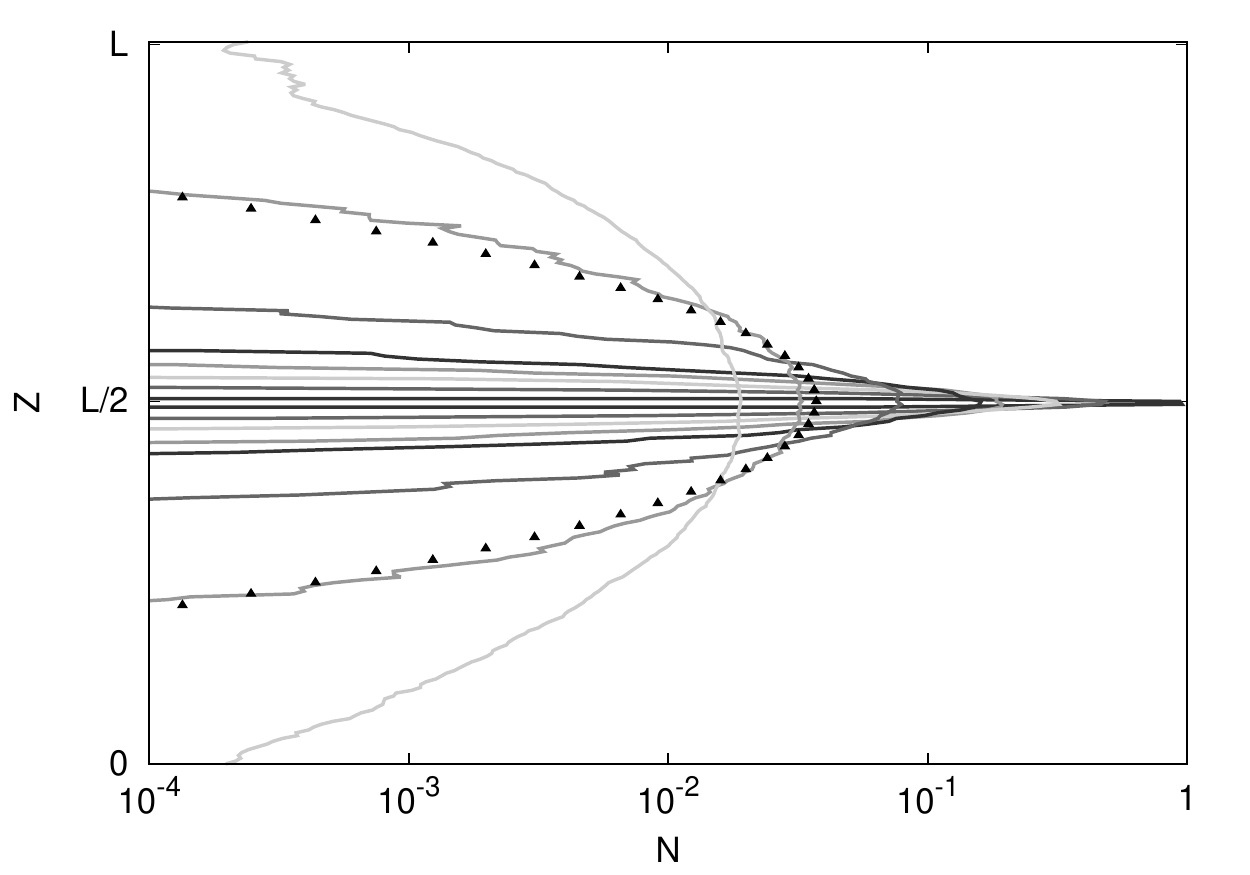}
  \caption{Plot of the normalized $z$-distribution of particles, $N_K(z)$, for different
    values of the force constant $K$ (or different values of
    $\sigma$). The curves corresponds to the values
    $\sigma/\eta = 38, 21, 9.1, 4.7, 3.48, 2.33, 1.41, 0.46$; the vertical width of the distribution  monotonically increases. The points show a normalized Gaussian fit for the curve with $\sigma/\eta = 21$. $\eta$ is the Kolmogorov length scale.
}
  \label{fig:z-hist-el}
\end{figure}

For unconfined particles we observe the classical homogeneous distribution on the whole volume,
while  in the extreme case of
particles bound to move on a plane, we observe a fractal-like distribution with a   dimension
considerably smaller than $2$. For each different value of the force constant we estimated the
probability density of finding a particle with a $z$ coordinate in the
range $[z;z+\delta z]$ computing an histogram, $N_K(z)$.
Figure \ref{fig:z-hist-el} shows the histograms for the
$z$ distribution of the particles. The distribution is well
approximated by a Gaussian with center at $z=z_0$ and variance
$\sigma^2$.

Each curve in Figure \ref{fig:z-hist-el} corresponds 
to a specific value of $\sigma$. There is a
one-to-one correspondence between the values of $\sigma$ and those of the
force constant $K$. Measuring $\sigma$ is thus an intuitive way
for quantitatively describing the strength of the confinement of the
particles around the central plane $z=z_0$.

\begin{figure}[!t]
  \begin{center}
    \includegraphics[width=\linewidth]{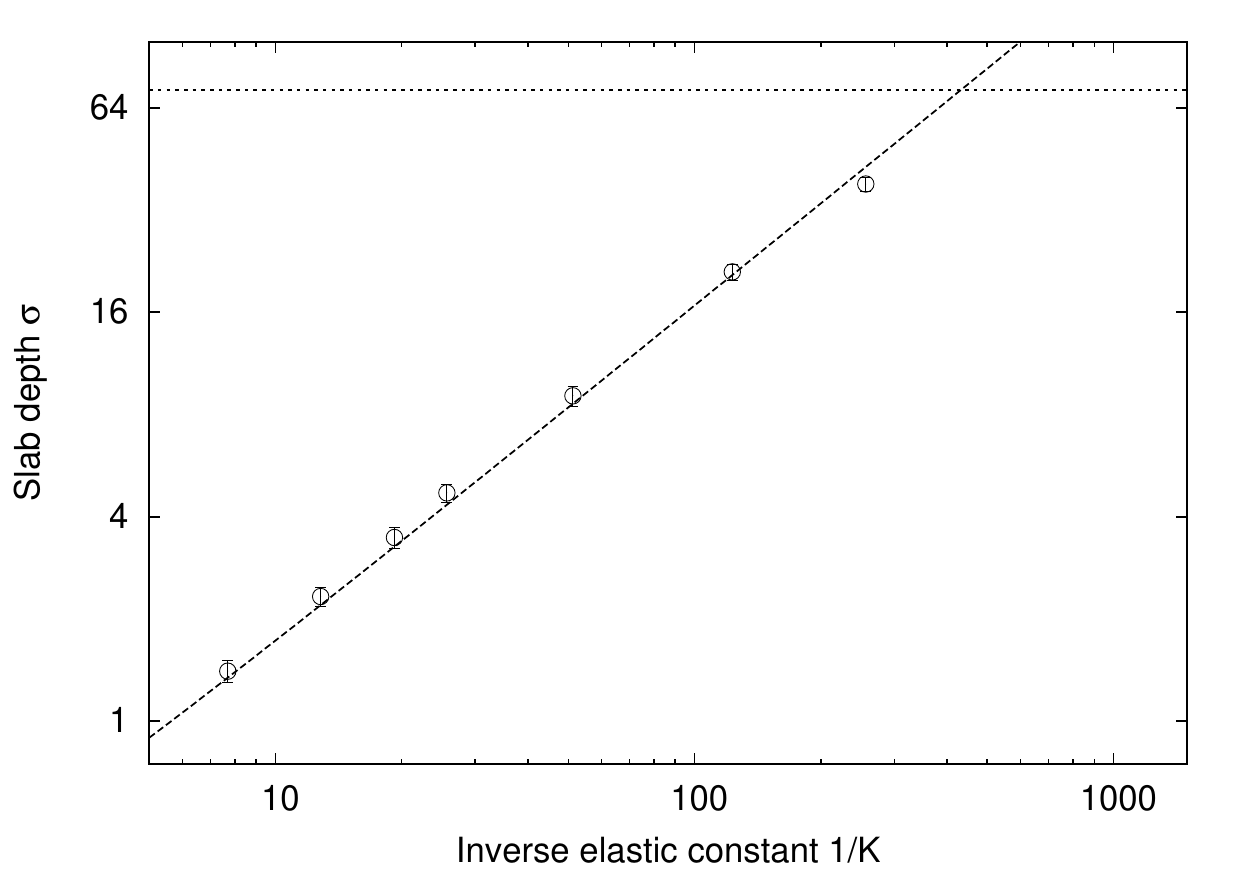}
  \end{center}  \caption{
Plot of the slab depth $\sigma$ versus the inverse of the
  force constant $1/K$. The horizontal dotted line 
 is the asymptotic value of $\sigma$ for $K \rightarrow 0$. The slab depth is 
 given in units of $\eta$, the Kolmogorov length scale. $1/K$ is 
 given in units of $\tau_{\eta}^{-1}$, where $\tau_{\eta}$ is the Kolmogorov time scale. 
 The diagonal dashed line is obtained from a linear least
squares fit.
}
  \label{fig:rmsd-vs-elc}
\end{figure}

For a linear restoring force, we expect $\sigma$ to be inversely
proportional to  $K$. Furthermore, it is possible
to calculate analytically the value of $\sigma$ in two limit cases:
unconfined particles or particles restricted to a plane.  In the limit of
particles strictly confined on a plane, i.e., $K \rightarrow \infty$,
$\sigma$ is exactly zero. In the limit of $K \rightarrow 0$, i.e.,  no
vertically constraining force, particles are freely advected over the
full domain following the underlying fluid motion (3D turbulent
diffusion). The particle density becomes homogeneous over the cubic
domain and $\sigma$ evolves to a constant value that is a fraction of $L$. 
Figure \ref{fig:rmsd-vs-elc} shows the relation between the dispersion
$\sigma$ and the force constant $K$. We see that, for large values
of $K$, $\sigma$ is proportional to ${1}/{K}$, as we expect. This proportionality 
should disappear when $K$ is decreased.
\subsubsection{2D and 3D compressibility}
\label{sec:met.compressibility}
We measured two different compressibilities for the system of
particles (the underlying flow is always incompressible): the 2D
compressibility based only on the horizontal velocity components and
the full 3D compressibility. In both cases, in order to obtain an
ensemble average, the compressibility has been calculated averaging 
on both particles and time. The error is then given by the standard deviation of
the mean.
\paragraph{2D compressibility}
\label{sec:2DCompressibility}
\noindent The 2D compressibility is defined as:
\begin{equation}
  C_{2D}=\frac{\left\langle ( \sum_i{\partial_i
      u_i})^2\right\rangle}{\left\langle \sum_{i,j}{ (\partial_i
      u_j)^2}\right\rangle} \, ,
  \label{eq:c2d}
\end{equation}
where $i,j=x,y$ and $u$ is the particle velocity.\\
In the limit $K \rightarrow 0$ we can calculate analytically the
value of $C_{2D}$. This limit corresponds to extracting the velocity data from a plane in a fully
3D flow field. For a 3D homogeneous and isotropic turbulent
flow (HIT), by simply substituting the relations between the velocity gradients inside Eq. \eqref{eq:c2d}, 
we find $C_{2D}^{HIT} =\frac{1}{6} \simeq 0.167$.

\begin{figure}[!t]
  \begin{center}
    \includegraphics[width=\linewidth]{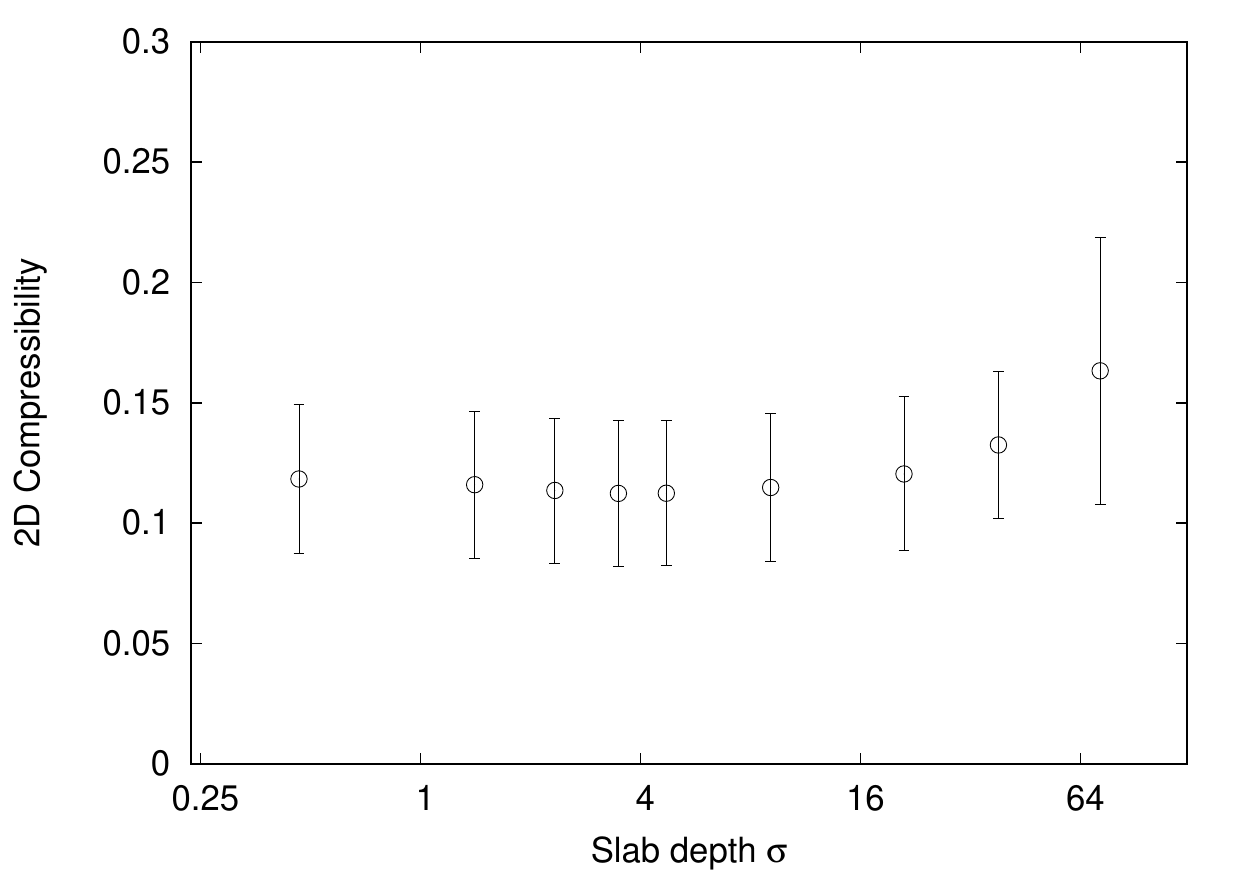}
  \end{center}
  \caption{Plot of the 2D compressibility $c_{2D}$ versus the slab thickness $\sigma$. Here, $\sigma$
    is given in units of the Kolmogorov length scale $\eta$. 
    Data points are obtained averaging on both time and particles. Errors are the standard deviation of the mean.
}
  \label{fig:2dcompr-el}
\end{figure}
Figure \ref{fig:2dcompr-el} shows the relation between the 2D
compressibility Eq. \eqref{eq:c2d} and particle dispersion $\sigma$
(that is inversely proportional to the force constant $K$).
In the case of unconfined particles we correctly recover $C_{2D} = C_{2D}^{HIT}$. 
Increasing the confinement, the 2D compressibility
has a value lower than $C_{2D}^{HIT}$. This is because the exact value $C_{2D}^{HIT}$ is calculated in an 
Eulerian framework, averaging the velocity field of the flow in the whole plane, while in
our case we use the ``Lagrangian'' velocity gradients, i.e., the
Eulerian velocity gradients measured at the positions of the particles
(thus not homogeneously distributed in a plane, due to the presence of
preferential concentration). Only without preferential concentration do we expect $C_{2D} = C_{2D}^{HIT}$.
\paragraph{3D compressibility}
\label{sec:3DCompressibility}
\noindent The 3D compressibility is defined as:
\begin{equation}
  C_{3D}=\frac{\left\langle ( \sum_i{\partial_i
      u_i})^2\right\rangle}{\left\langle \sum_{i,j}{ (\partial_i
      u_j)^2}\right\rangle} \, ,
  \label{eq:c3D}
\end{equation}
where $i,j=x,y,z$ and $u$ is the particle velocity.\\
Substituting the equations of motion~\eqref{eq:eq_motion_particles} in
Eq. \eqref{eq:c3D} we obtain:
\begin{eqnarray}
  \label{eq:c3D_bis}
  C_{3D}=\frac{\left\langle (\partial_x v_x + \partial_y v_y +
    \partial_z v_z - K)^2\right\rangle}{\left\langle \left[
      \sum_{i,j}{ (\partial_i v_j)^2} \right] - 2 K \partial_z
    v_z + K^2\right\rangle} = \nonumber \\
  = \frac{K^2}{\left\langle
    \sum_{i,j}{ (\partial_i v_j)^2} \right\rangle + K^2} \, ,
\end{eqnarray}
where we explicitly used the incompressibility of the 3D flow field
and the fact that $\langle \partial_z v_z\rangle = 0$. So we
expect $C_{3D}=0$ in the $K \rightarrow 0 $ limit, and $C_{3D}=1$ in
the $K \rightarrow \infty$ limit.

\begin{figure}[!t]
  \centering \includegraphics[width=\linewidth]{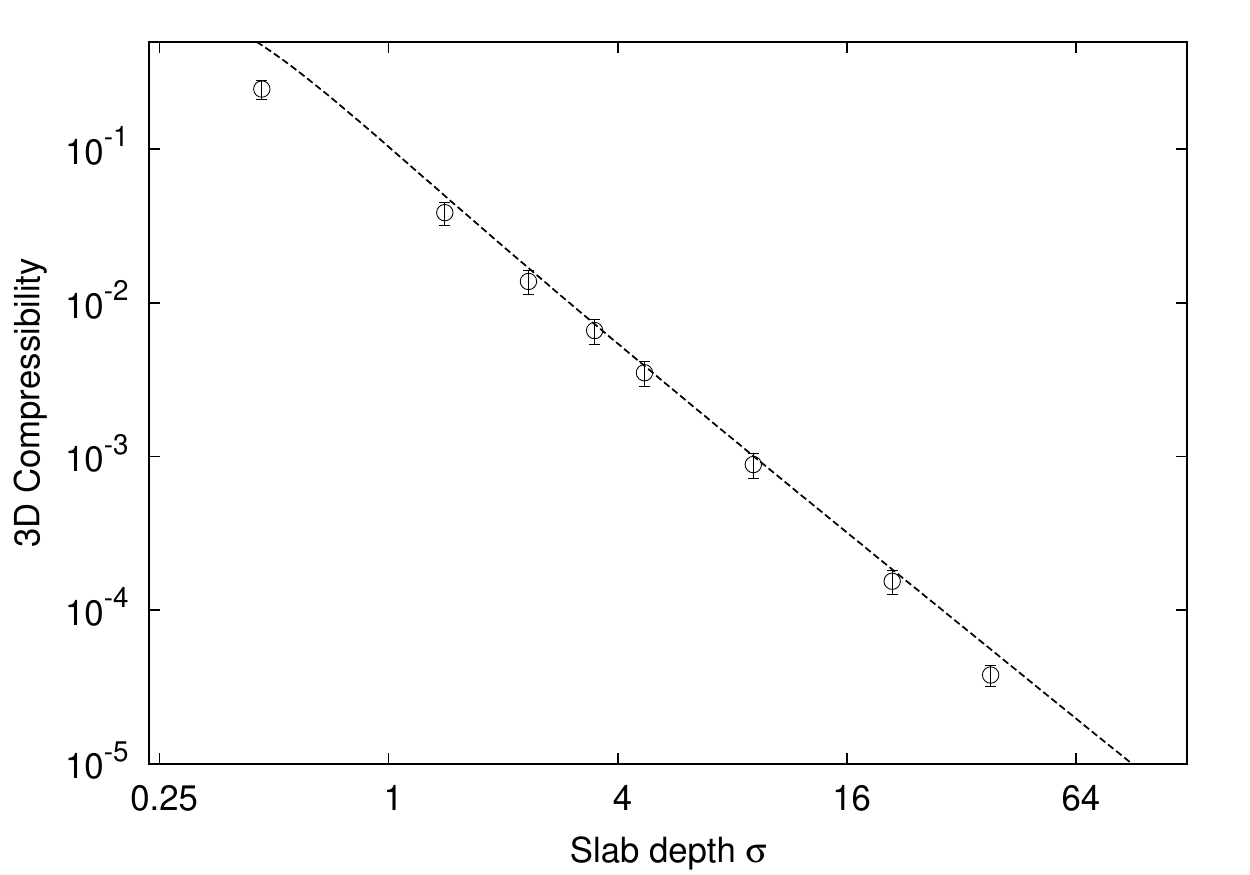}
  \caption{Log-log plot of the 3D compressibility $C_{3D}$ versus
    the slab thickness $\sigma$. Here $\sigma$ is given in units of the
    Kolmogorov length scale $\eta$. The dashed line is a fit of Eq.~\eqref{eq:c3D_bis}.
}
  \label{fig:3Dcompr-el}
\end{figure}

Figure \ref{fig:3Dcompr-el} shows the relation between the 3D
compressibility~\eqref{eq:c3D} and the particle dispersion $\sigma$
(that is inversely proportional to the force constant $K$). As
expected, $C_{3D}$ is zero for  tracers, and increases
monotonically with the confinement strength.

\subsubsection{Accumulation of particles and pair correlations in space}
\label{sec:met.correlation}
Another important way of characterizing the system is to look at the
distribution  of particles at small scales, i.e., the tendency of particles to
inhomogeneously concentrate in space. This behavior is the result of
the interplay between the motion induced by the underlying fluid and
the presence of a vertically confining force on the particles.
Because of the inhomogeneity on the vertical direction, we decided to characterize the  spatial distribution centering the analysis on 
 a small volume around the central equilibrium plane:
\begin{equation}
  A = \left\{ z_{i} \in \left[ L/2 - \Delta z ; L/2 + \Delta z \right] \right\} \, ,
  \label{eq:pdilimits}
\end{equation}
with $\Delta z \simeq 0.2\eta$.
In this way, measurements on horizontal scales larger (smaller) than $\Delta z$  will be mainly two-dimensional (three dimensional).

We defined a  pair distribution integral, $P_2(r)$  as follows: having defined the set $A$ of all particles falling in the central volume of vertical width $\Delta z$, one counts how many pairs with a relative distance $\le r$ can be formed with a particle in $A$ and another particle anywhere in the volume. Formally:
\begin{equation}
  P_2(r)=\sum_{i \in A}\sum^{N_{p}}_{j=1}\Theta(r - \left|
  \mathbf{x}_{i} - \mathbf{x}_{j} \right|) \, ,
  \label{eq:pdi}
\end{equation}
where $\Theta(x)$ is the Heaviside step function and
$\mathbf{x}_{i}$, $\mathbf{x}_{j}$ are the particle coordinates. 
If one had chosen $A$ equal to
the whole system of particles then one would obtain  the classical correlation dimension \cite{grassberger1983measuring}.\\ 

In order to quantify the  scale-by-scale clustering properties it is useful to introduce the local scaling-exponent
\begin{equation}
  \zeta (r) = \frac{d \log \left( P_2(r') \right)}{d \log \left(r'\right)} \Big\vert_{r'=r} \, .
  \label{eq:derlog}
\end{equation}

In the limit $r \rightarrow 0$ one expects that the scaling exponent 
recovers the definition of  correlation dimension of the
particle distribution.

This local
scaling exponent expresses how the number of pairs scales with the
distance $r$, for the particles in the innermost part of the
layer. 

\begin{figure}[!t]
  \centering
  \includegraphics[width=\linewidth]{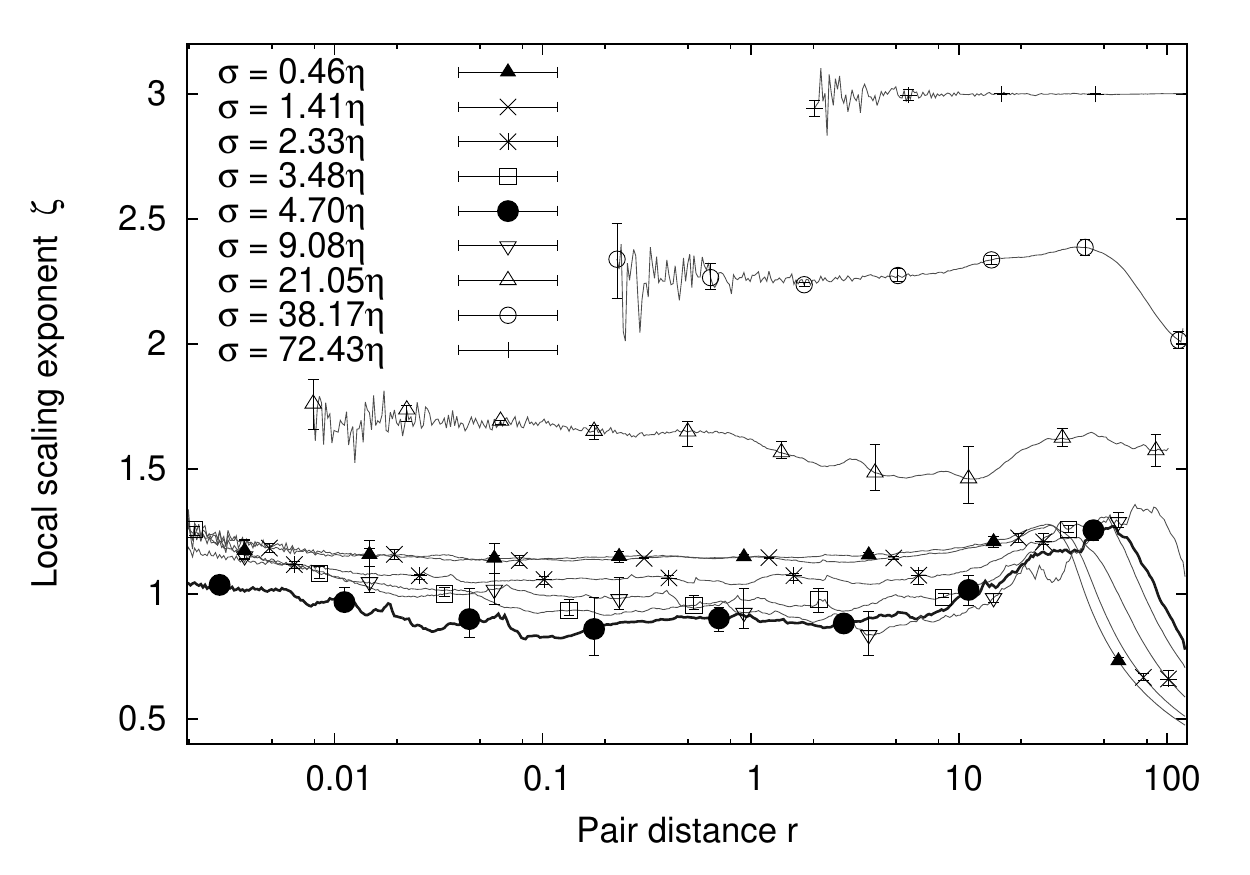}
  \caption{Log-log plot of the local scaling exponents $\zeta(r)$
    versus the radial distance of the pairs, for different values of
    the effective slab thickness $\sigma$. 
 	Continuous
    lines represent all the available data; the purpose of the
    symbols is to help the reader to distinguish the different curves.
    Both the slab thickness $\sigma$ and the pair distance $r$ are
    given in units of the Kolmogorov length scale $\eta$. Only a few indicative
    error bars are shown, in order to keep the Figure clear. The error 
    bars have been estimated by comparing results between two different
	subsets of the whole statistics. The curve corresponding to $\sigma =
    4.70\eta$ is emphasized to stress the fact that the minimum in
    fractal dimension does not correspond to the minimum in slab
    thickness.}
  \label{fig:exponent-el-vol}
\end{figure}

\begin{figure}[!t]
  \centering
  \includegraphics[width=\linewidth]{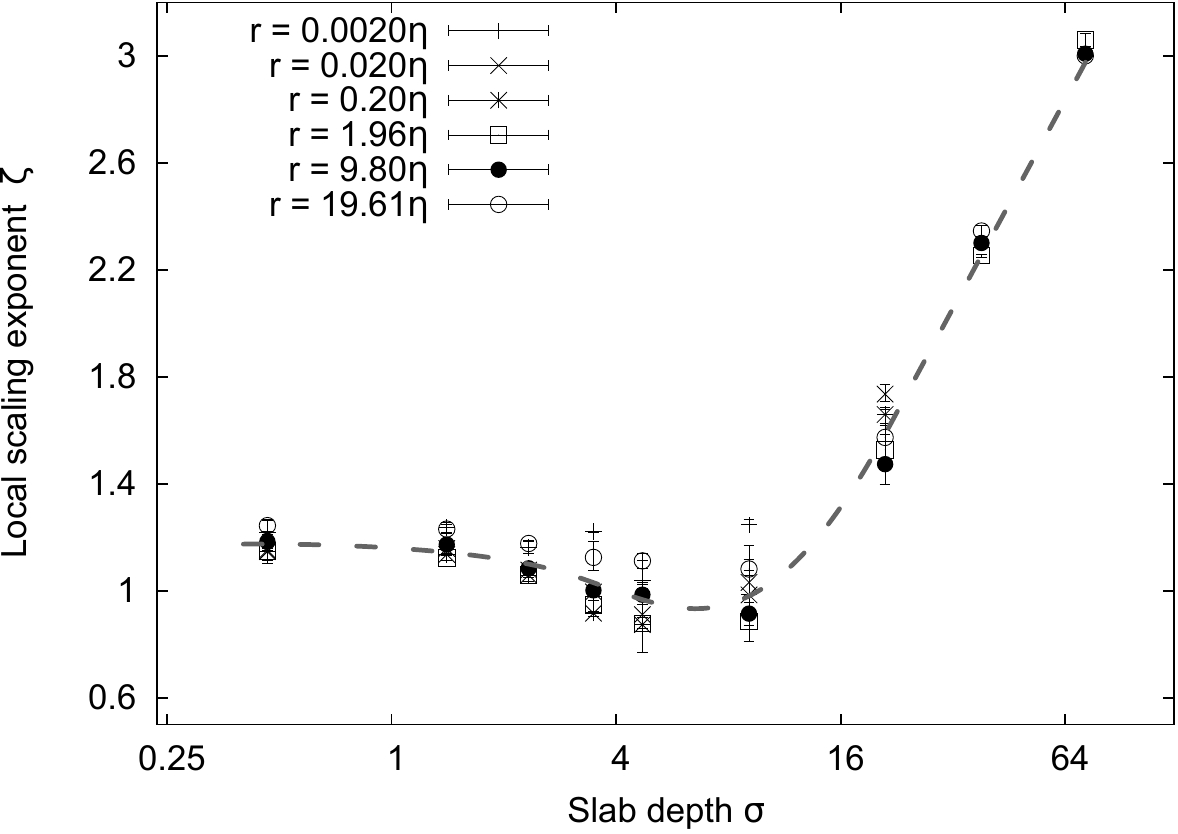}
  \caption{Plot of the value of the power-law exponent $\zeta(r)$
    versus $\sigma$ for different values of $r$.  We can see that the
    curves in the range $0.02 < r/\eta< 19.6$ exhibit
    non-monotonicity, and a minimum around $r \simeq 5\eta$. Points
    are taken intersecting the curves in Figure
    \ref{fig:exponent-el-vol} with lines $r={\rm{constant}}$. 
	 Both the slab thickness $\sigma$ and the pair distance $r$ are
    given in units of the Kolmogorov length scale $\eta$. The error
    estimation procedure is detailed at the end of Sec. \ref{sec:met.correlation}. 
    The dashed line is merely a guide for the eye.}
  \label{fig:exponent_vs_rmsd_vol}
\end{figure}

Figure \ref{fig:exponent-el-vol} shows the local
scaling exponent $\zeta(r)$ versus the radial distance of the pairs, where the error bars have been estimated by comparing results between two different
subsets of the whole statistics. 
Figure \ref{fig:exponent_vs_rmsd_vol} show the local scaling exponents
$\zeta(r)$ at different values of the distance $r$ versus the particle
dispersion $\sigma$. This figure confirms the results of Fig. \ref{fig:exponent-el-vol}: there exists a minimum in the
exponent (corresponding to a maximum in the accumulation of particles)
for $\sigma \simeq 5\eta$, at least in a range of scales $0.02 <
r/\eta< 19.6$. For $\sigma\gtrsim 5\eta$ the exponent grows, corresponding to
a decrease in clustering because of the reduced confinement effects. On the other hand,
if $\sigma$ is decreased below $\eta$ the exponent
stays constant, since the particles are already constrained to be very close to the central plane.

Let us notice that using the correlation integral as introduced in  \cite{grassberger1983measuring} to analyze the particle distribution leads to 
an undesirable effect for our set-up: centering the spheres on peripheral particles (far from the central plane),
gives a lower number of pairs inside a given sphere of radius $r$ because of the vertical non-homogeneous distribution.
Using our definition to  measure the pair distribution integral
$P_2(r)$  corresponds to measuring the original three-dimensional distribution  in such a way that the ``central'' particles have a larger
weight with respect to the peripheral ones, and the effect of the
inhomogeneity in the $z$ direction is less pronounced.
The value
$\Delta z \simeq 0.2\eta$ has been chosen because it allows us to
shift the abrupt drop in the scaling exponent ( visible in Fig.
\ref{fig:exponent-el-vol}) at large scales, leaving a cleaner
power-law behavior at the scales we are interested in. 
Increasing
$\Delta z$ would shift the drop in the scaling exponent at smaller
length scales.
\section{Conclusions}
In this study we have investigated the dynamics of a system of
particles suspended in a turbulent flow and vertically constrained to
evolve within an horizontal slab with a certain thickness depending on
an effective linear restoring force.  In particular, we quantified the
effective compressibility of the particle distribution and particle
accumulation varying the degree of confinement.

Using DNS we have studied a simplified model in which particles are
suspended in a 3D, isotropic and homogeneous, turbulent flow. These
particles are passive, point-like and confined only in the vertical
direction by means of a restoring force. We studied different
situations, varying the thickness of the slab, in order to analyze the
characteristics of the particle suspension in a range of conditions
from having full accessibility to the 3D incompressible flow, to a 2D
slice of a 3D incompressible flow. The particle distribution shows a
certain degree of compressibility, except for $K\rightarrow 0$.

We have also shown that there exists a particular -optimal- value of
the effective slab thickness $\sigma$ (or, equivalently, of the
force constant $K$) that maximizes the accumulation of the particles
(minimizing the fractal dimension of the system). This happens when
the depth of the horizontal slab is of the order of a few Kolmogorov
length scales $\eta$ ($\sigma \sim 5\eta$).
Let us point out that viscous effects are known to be important 
up to $5-10\eta$ in turbulent
flows, meaning that the maximum particle accumulation is achieved when 
their vertical displacement is bounded to be no larger than the size of viscous eddies.

We want to stress how our model, though simple, could be important  towards the quantitative understanding of the
phenomenology of passive, point-like entities in a turbulent marine
thin layer. Indeed, our model captures the generic features associated
with the presence of a vertical localization, irrespective of the
biological or physical reason beyond its production and its stability.
Thin phytoplankton layers are always much thicker than the size of
individual cells and for this reason one may question how the
discussed confinement may be relevant at all to plankton population
dynamics. 
Here it must be stressed that what is important is the relation
between time scales, and not length scales, in the system. Indeed the
typical generation time for plankton and bacteria in a marine
environment is well within the inertial range and such that over a
generation the cell has been transported to distances much larger than
the vertical confinement.
Clearly real marine conditions are complicated by many more phenomena
that we have not included in our model and the real plankton dynamics
shows phenomena that the simplified model discussed here cannot
incorporate.
The improvement of our model in order to apply it to more complex
situations and to the modeling of systems more similar to real-life
plankton particles in oceans is a challenge for future studies. For
example, an obvious follow-up to this investigation would be to
simulate inertial particles instead of passive tracers, integrating
more physically accurate equations of motion, while keeping a simple linear
restoring force for the vertical confinement.\\

\section*{ACKNOWLEDGMENTS}
This work is part of the Research Programs No. 11PR2841 
and No. FP112 of the Foundation for Fundamental Research
on Matter, which is part of the Netherlands Organisation for
Scientific Research. The work of L.B. and M.D.P was partially
funded by European Research Council Grant No. 339032.
We acknowledge support from the European Cooperation
in Science and Technology (COST) Actions MP0806 and
MP1305.

\bibliographystyle{unsrt}
\bibliography{bibliography}
\end{document}